\documentclass{article}
\usepackage{float}
\usepackage{authblk}                    %authors
\usepackage{graphicx,epsfig,epstopdf}   %figures
\usepackage{booktabs}                   %tables
\usepackage{amsmath,amsthm,amssymb}     %math
\usepackage{color}                      %color
\usepackage{cite}                       %citation
\usepackage{lipsum}
\DeclareUnicodeCharacter{2212}{-}

\graphicspath{ {figures/} }

\usepackage[paperwidth = 17cm,          %size
paperheight = 24cm,
left = 1.75cm,
right = 1.75cm,
top = 2cm,
bottom = 2cm]{geometry}

\usepackage[
    bookmarks         = true,%     % Signets
    bookmarksnumbered = true,%     % Signets numérotés%
    bookmarksopen     = true,%     % Signets ouverts
    colorlinks        = true,%     % Liens en couleur : true ou false
    linktocpage       = true,%
    urlcolor          = blue,%     % Couleur des liens externes
    linkcolor         = blue,%     % Couleur des liens internes
    citecolor         = red,%      % Couleur des citations
    ]{hyperref}%

\definecolor{coolblack}{rgb}{0.0, 0.18, 0.39}

\begin{document}

\title{\textbf{Study of gravitational waves from phase transitions in three-component dark matter}}
\author[1]{Mohammad Hossein Rahimi Abkenar\thanks{rahimi.mh@fs.lu.ac.ir}}
\author[1]{Ahmad Mohamadnejad\thanks{mohamadnejad.a@lu.ac.ir}}
\author[1]{Reza Sepahvand\thanks{sepahvand.r@lu.ac.ir}}
\affil[1]{Department of Physics, Lorestan University,
Khorramabad, Iran}

\date{\today}  
\maketitle

\begin{abstract}
This paper studies gravitational waves in a dark matter model composed of three types of particles with distinct spins, along with a scalar field $\phi$ that mediates interactions between Standard Model particles and dark matter. It discusses the electroweak phase transition following the Big Bang, during which all particles are initially massless due to the inactive Higgs mechanism. As temperature decreases, the effective potential reaches zero at two points, leading to two minima at the critical temperature ($T_c$), and eventually to a true vacuum state. The formation of new vacuum bubbles, where electroweak symmetry is broken and particles acquire mass, generates gravitational waves as these bubbles interact with the fabric of space-time. The paper derives the gravitational wave frequency and detection range based on the model's parameters, aligning with observational data from the Planck satellite and detection thresholds from PandaX-4T and XENONnT for some parameter points. It concludes by comparing the predicted background gravitational wave density with the sensitivities of LISA, BBO and $\mu$-Ares detectors.

\end{abstract}

%\newpage
%\noindent\hrulefill
%\tableofcontents
%\noindent\hrulefill

\numberwithin{equation}{section}

\hrule height 0.5pt

\tableofcontents 

\vspace{10pt}
\hrule height 0.5pt
\section{Introduction} \label{sec1}
After the Big Bang, all particles, including dark matter particles undergoes an electroweak phase transition at a certain temperature. We present a dark matter model composed of three types of particles, each with distinct spins, along with a scalar field $\phi$ that serves as an intermediary between Standard Model particles and dark matter, capable of assuming the vacuum expectation value. Prior to electroweak phase transition, three-component dark matter particles and all SM particles are massless, because the Higgs mechanism has not yet activated. Before the electroweak phase transition, the minimum of effective potential at a certain value of the scalar field $\phi$ is equal to zero, which indicates the initial vacuum before the phase transition. As the temperature decreases, the effective potential becomes zero at two distinct points of the scalar field $\phi$, resulting in two minima corresponding to the critical temperature ($T_c$). As the temperature continues to drop, the potential reaches a distinct negative value, leading to a single minimum known as the "true vacuum". Cosmic phase transitions happen when the temperature decreases below a critical level, resulting in the Universe transitioning from a symmetric phase to one characterized by broken symmetry. This temperature, at which the cores of true vacuum bubbles begin to form in space, is known as the nucleation temperature ($T_n$). In fact, with the breaking of symmetry, the scalar field ($\phi$) acquires a nonzero vacuum expectation value (VEV), and all particles gain mass through the Higgs mechanism. New vacuum bubbles form spherically and move through space, replacing the vacuum before the phase transition. Inside the bubbles, electroweak symmetry is broken, and the Higgs mechanism is activated. In fact, inside the bubbles, dark matter and Standard Model particles have mass, whereas outside these bubbles, they are massless \cite{Zurek:2008qg,Biswas:2013nn,Abkenar:2024ket,YaserAyazi:2018lrv,Guth:1981uk,Steinhardt:1981ct,Witten:1980ez,Steinhardt:1980wx,Witten:1984rs,YaserAyazi:2019caf,Bertone:2019irm,Croon:2020cgk,Lu:2025vif,Zhou:2025zzz,Benincasa:2024pfs,Allahverdi:2024ofe,Ghorbani:2024twk,Ramsey-Musolf:2024zex,Cirelli:2024ssz,Xu:2023lkf}.  

The motion of spherically shaped new vacuum bubbles (true vacuum) within the hot plasma resulting from the Big Bang leads to expansion and oscillation in the space-time fabric due to the distribution of mass in space. As bubble walls collide, they induce perturbations in the surrounding space-time. This occurs at the percolation temperature ($T_p$) and leads to the emission of gravitational waves that travel through space. At the percolation temperature, bubbles occupy roughly 30\% of the Universe’s volume \cite{Athron:2022mmm}. These gravitational waves were produced in the early universe, and their strength has diminished significantly over time and the expansion of the universe, ultimately transforming into cosmic background gravitational waves. In this paper, we derive the GW frequency and range for the detection of these background gravitational waves, considering the assumption of the existence of multi-component dark matter. The first order electroweak phase transition can be acquired by various dark matter models \cite{Hall:2019rld,Hall:2021zsk,Hall:2019ank,Chala:2016ykx,Baldes:2017rcu,Flauger:2017ged,Chao:2017vrq,Han:2020ekm,Deng:2020dnf,Kannike:2019wsn,Croon:2024mde,Croon:2022tmr,Athron:2023rfq,Athron:2023xlk}.  

In this article, we first present the Lagrangian and its dependent constraints for our dark matter model based on WIMPs. In our model, a pair is considered for the Higgs particle, and its mass ($M_{H_2}$) is determined by the Gildener–Weinberg mechanism. Next, we select points from our model's parameter space where their relic density aligns with the Planck satellite report and their rescaled DM-nucleon cross sections exceed the PandaX-4T and XENONnT detection threshold. We then calculate the effective potential for these points and, using the Euclidean action, determine the necessary parameters for background gravitational waves, which we substitute into the relevant equations. Finally, We determine the detection range for the background gravitational wave density using a three-component dark matter model and compare it to the sensitivities of the LISA, BBO and $\mu$-Ares gravitational wave detectors \cite{PandaX-4T:2021bab,Planck:2018vyg,Masoumi:2017trx,Apreda:2001us,Ellis:2018mja,Crowder:2005nr,Caprini:2019egz,LISA:2017pwj,Sesana:2019vho}.

The paper is organized as follows: Section \ref{sec2} introduces the Lagrangian of our model, while Section \ref{sec3} examines the electroweak phase transition via the one-loop effective potential. In Section \ref{sec4}, we present our results, considering the gravitational wave spectrum using nine selected points outlined in Table \ref{tableOmega}. We conclude in Section \ref{sec5}.

\section{The model} \label{sec2}
\subsection{Lagrangian of model} \label{sec2-1}
In this section, we use a model for dark matter that we previously introduced in our earlier article \cite{Abkenar:2024ket}. We proposed a model featuring a scalar field \( S \), two spinor fields \( \psi^{1,2} \) (including right-handed \( \psi^{1,2}_R \) and left-handed \( \psi^{1,2}_L \)), and a vector field \( V_\mu \) as dark matter (DM), along with a complex scalar field \( \phi \) acting as an intermediate particle. All these fields are singlets under the Standard Model gauge group. We introduce a discrete symmetry under which the new fields transform as follows:

\begin{equation} \label{z2symmetry}
\phi \rightarrow \phi^{*}, \quad S \rightarrow -S, \, \quad V_{\mu} \rightarrow - V_{\mu}, \, \quad \psi_L^1 \rightarrow -\psi_L^2  \quad \text{and}  \, \quad \psi_R^1 \rightarrow -\psi_R^2.
\end{equation}

All SM particles are even and singlets under the dark gauge symmetry. The local gauge transformation for the new fields is as follows:

\begin{align} \label{invariant}
& S \rightarrow e^{i Q_S \alpha(x)} S, \nonumber \\
& \psi_L^a \rightarrow e^{i Q_L^a \alpha(x)} \psi_L^a, \nonumber \\
& \psi_R^a \rightarrow e^{i Q_R^a \alpha(x)} \psi_R^a, \nonumber \\
& \phi \rightarrow e^{i Q_{\phi} \alpha(x)} \phi, \nonumber \\
& V_{\mu} \rightarrow V_{\mu} - \frac{1}{g_v} \partial_{\mu}{\alpha(x)}.
\end{align}

In our model, the scalar field \( \phi \) and the spinor fields \( \psi^{1,2}_R \) and \( \psi^{1,2}_L \) carry charge under a dark \( U_D(1) \) gauge symmetry, with the vector field \( V_{\mu} \) acting as the gauge field (see Table~\ref{charge}).

\begin{table}[H] 
\centering
%\vspace{1cm}
\parbox{8.5cm}{\caption{The charges associated with the particles in the dark sector with respect to the newly introduced \( U_D(1) \) symmetry. \label{charge}}}

\vspace{7pt}
\begin{tabular}{ l  l  l  l  l  l  l  l}
\hline
field&$\phi$&$S$&$V_{\mu}$&$\psi_L^1$&$\psi_R^1$&$\psi_L^2$&$\psi_R^2$ \vspace{2pt} \\
\hline
dark charge ($ Q $)&$1$&0&0&$\frac{1}{2}$&$-\frac{1}{2}$&$-\frac{1}{2}$&$\frac{1}{2}$ \vspace{2pt} \\
\hline
\end{tabular}
\end{table}

We present the Lagrangian, invariant under the local gauge transformation (\ref{invariant}) and incorporating renormalizable interactions, as follows:

\begin{align}  \label{tlagrangian}
{\cal L} ={\cal L}_{SM}+\frac{1}{2}(\partial_{\mu} S)(\partial^{\mu} S) + (D_{\mu} \phi)^{*} (D^{\mu} \phi) -\frac{1}{4} V_{\mu \nu} V^{\mu \nu}- V(H,S,\phi) \nonumber \\
+\sum_{a=1}^2 \left(  i\bar\psi_L^a  \gamma^{\mu}D_{\mu}\psi_L^a+ i \bar\psi_R^a \gamma^{\mu}D_{\mu}\psi_R^a\right)
-g_{\phi,1} \phi \bar\psi_L^1 \psi_R^1 - g_{\phi,2} \phi^{*} \bar\psi_L^2 \psi_R^2 + {\text{H.C.}} ,
\end{align}
where $ {\cal L} _{SM} $ is the SM Lagrangian without the Higgs potential term. The covariant derivative and field strength tensor are $D_{\mu}= (\partial_{\mu} + i Q g_{v} V_{\mu})$ and $V_{\mu \nu}= \partial_{\mu} V_{\nu} - \partial_{\nu} V_{\mu} $, respectively. From (\ref{z2symmetry}), we have \( g_{\phi,1}=g_{\phi,2}=g_{\phi} \), leading to $M_{\psi_1}=M_{\psi_2}=M_{\psi}$.  We introduce gauge invariant potential $V(H,\phi,S)$ as follows:

\begin{align} \label{vlagrangian}
V(H,\phi,S) = & \, \lambda_{H} (H^{\dagger}H)^{2} + \lambda_{\phi} (\phi^{*}\phi)^{2}  
+ \lambda_{H \phi} (H^{\dagger}H)(\phi^{*}\phi) \nonumber \\
&+ \frac{1}{2} \lambda_{H S} (H^{\dagger}H)S^2
+ \frac{1}{2} \lambda_{\phi S} (\phi^{*}\phi)S^2 + \frac{1}{4} \lambda_{S} S^4.
\end{align}

Since all the particles introduced as dark matter are odd, no interaction occurs between them, leading to an accidental symmetry that allows dark matter to persist. The scalar field \( \phi \) can acquire VEVs that break the \( U_D(1) \) symmetry, while the Higgs field \( H \) can obtain VEVs that break electroweak symmetry. We can rewrite the scalar and Higgs fields in unitary gauge as follows:

\begin{equation} \label{gauge}
H = \frac{1}{\sqrt{2}} \begin{pmatrix}
0 \\ h_{1} \end{pmatrix} \, \, \, and \, \, \, \phi = \frac{1}{\sqrt{2}} h_{2}.
\end{equation}

Here, \( h_{1} \) and \( h_{2} \) are real scalar fields that can acquire vacuum expectation values. By substituting Eqs. (\ref{gauge}) into Eq. (\ref{vlagrangian}), the tree-level potential can be rewritten as follows:

\begin{equation} \label{vtree}
V^{tree} = \frac{1}{4} \lambda_{H} h_{1}^{4} + \frac{1}{4} \lambda_{\phi} h_{2}^{4} + \frac{1}{4} \lambda_{\phi H} h_{1}^{2} h_{2}^{2} + \frac{1}{4} \lambda_{H S} h_{1}^2 s^2 + \frac{1}{4} \lambda_{\phi S} h_{2}^2 s^2 + \frac{1}{4} \lambda_{s} s^4  . 
\end{equation}

The local minimum of the tree-level potential (\ref{vtree}) at \( \langle h_{1} \rangle = \nu_{1} \) and \( \langle h_{2} \rangle = \nu_{2} \) defines the VEVs of the fields, leading to the following condition:

\begin{align} 
& \lambda_{\phi H} < 0 \;\; , \;\; \lambda_H > 0\;\; , \;\; \lambda_\phi > 0 \nonumber \\
& \frac{\nu_1^{2}}{\nu_2^{2}}= - \frac{\lambda_{\phi H}}{2 \lambda_H}\;\; , \;\; \frac{\nu_2^{2}}{\nu_1^{2}}= - \frac{\lambda_{\phi H}}{2 \lambda_\phi}, \nonumber \\
& \;\;\;\;\;\;\;\;\;\;\;\; \lambda_{\phi H}^2=4 \lambda_H \lambda_\phi. \label{signs}
\end{align}

In the flat direction of field space, the tree-level potential reaches its minimum, along which \( V_{\text{tree}}(\nu_{1}, \nu_{2}, 0) = 0 \). Now, we substitute \( h_{1} \to \nu_{1} + h_{1} \) and \( h_{2} \to \nu_{2} + h_{2} \), mixing \( h_{1} \) and \( h_{2} \). The mass eigenstates, $H_1$ and $H_2$, can be obtained from the following rotation:

\begin{equation} \label{rotationMatrix}
\begin{bmatrix}
H_1 \\ 
H_2
\end{bmatrix}=\begin{bmatrix}
\cos \alpha & - \sin \alpha \\
\sin \alpha & \;\;\; \cos \alpha
\end{bmatrix} \begin{bmatrix}
h_1 \\ h_2
\end{bmatrix}.
\end{equation}

\begin{figure} [H] 
\centering
\centerline{\hspace{0cm}\fbox{\epsfig{figure=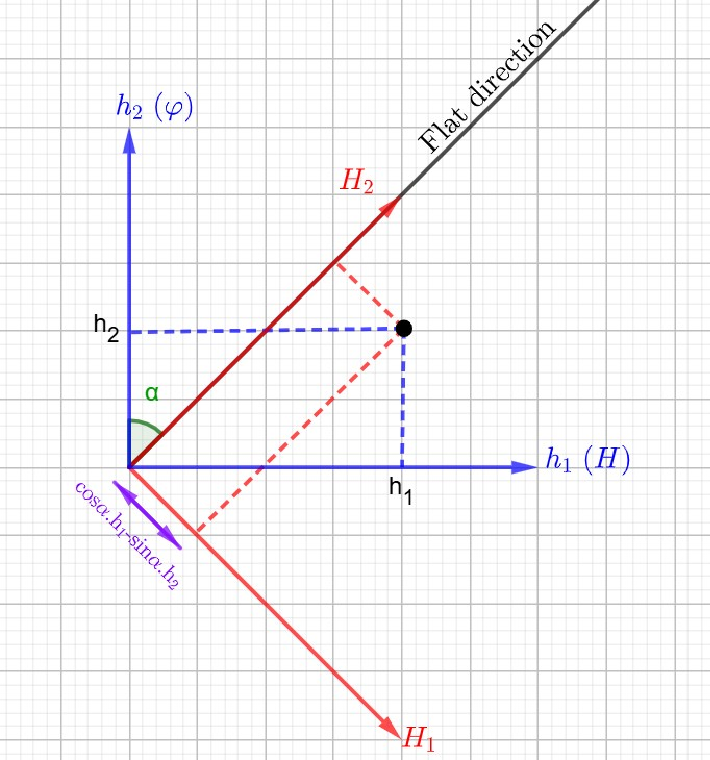,width=6cm}}}
\parbox{6.5cm}{\caption{The relative positions of the Higgs and scalar fields as indicated in (\ref{rotationMatrix}). \label{FlatPic}}} 
\end{figure}

The rotation of the fields in (\ref{rotationMatrix}), the flat direction position and ($\alpha$) angle are shown in figure \ref{FlatPic}. The mass of $H_1$ and $H_2$ are determined by substituting (\ref{rotationMatrix}) into (\ref{vtree}) and calculating $M_{H_i}^2 = \frac{\partial^2 V^{\text{tree}}}{\partial H_i^2}$, as follows: 

\begin{align} 
& M_{H_1}^2 =-H_2^2 \lambda_{\phi H}, \label{FirstMH1} \\
& M_{H_2}^2 =-H_1^2 \lambda_{\phi H}. \label{FirstMH2} 
\end{align}
Equation (\ref{FirstMH1}) and (\ref{FirstMH2}) are derived from the flat direction situation. Equation (\ref{FirstMH2}) shows that the mass of $H_2$ is influenced by the $H_1$ field, so in the flat direction where $H_1=0$, as illustrated in figure \ref{FlatPic}, the mass of $H_2$ is zero. The $H_1$ field is perpendicular to the flat direction and is identified as the SM-like Higgs observed at the LHC, leading us to consider $M_{H_1}=125 \; GeV$ \cite{CMS:2012qbp}. Figure \ref{FlatPic} shows that at every point along the flat direction line, the value of $H_2$ equals the hypotenuse.

\noindent Consequently, the average $<H_2>$ in the flat direction is calculated as follows:

\begin{equation} \label{hypotenuse}
<H_2>=\sqrt{\nu_1^2+\nu_2^2}=\nu.
\end{equation}
On the other hand, from the Standard Model, we know that $\nu_1 = 246 \; GeV$. As mentioned before, the field $H_2$ is aligned along the flat direction; thus, its tree-level mass is zero. However, the inclusion of one-loop corrections via Gildener–Weinberg mechanism leads to a non-zero value along the flat direction, yielding a minimum of the potential, which provides the following mass to $H_2$:

\begin{equation}  \label{MH2}
M_{H_{2}}^{2} = \frac{1}{8 \pi^{2} \nu^{2}} \left( M_{H_{1}}^{4} + 6  M_{W}^{4} + 3  M_{Z}^{4} + 3  M_{V}^{4} + M_{S}^{4} - 8 M_{\psi}^{4} - 12 M_{t}^{4}   \right),
\end{equation}
where $M_{W,Z,V,S,\psi,t}$ being the masses for W and Z gauge bosons, vector DM, scalar DM, spinor DM and top quark, respectively after symmetry breaking. The model's dependent constraints are obtained as follows:

\begin{align}
& \sin \alpha =  \frac{\nu_{1}}{\sqrt{\nu_{1}^{2}+\nu_{2}^{2}}}, \;\; g_{V}=\frac{M_{V}}{\nu_{2}}, \;\; g_{\phi}=\frac{\sqrt{2} M_{\psi}}{\nu_{2}},  \nonumber \\
& \lambda_{H} = \frac{M_{H_{1}}^{2} cos^2\alpha}{2 \nu_{1}^{2}}, \;\; \lambda_{\phi} = \frac{M_{H_{1}}^{2} sin^2\alpha}{2 \nu_{2}^{2}}, \nonumber  \\
& \lambda_{\phi H} =  - \frac{M_{H_{1}}^{2} \sin \alpha \cos \alpha}{ \nu_{1}\nu_{2}}, \;\; \lambda_{H S} = \frac{2 M_{S}^{2} - \lambda_{\phi S} \nu_{2}^2}{\nu_{1}^2}. \label{constrins}
\end{align}

There are six free parameters which we choose them as $M_S$, $M_V$, $M_\psi$, $\lambda_{\phi s}$, $\nu_2$ and $\lambda_s$. Conversely, $\lambda_s$ is irrelevant to the DM phenomenology and phase transition studied in this paper, leaving us with only four free parameters.

\subsection{Appropriate points} \label{sec2-2}
In this section, we use the micrOMEGAs to select nine points from parameter space of our model that achieve a relic density between 0.96 and 0.14, that some of these points have DM–nucleon cross sections that exceed the detection thresholds of the PandaX-4T and XENONnT detectors \cite{Hur:2007ur,PandaX-4T:2021bab}. For $M_{DM}>40\;GeV$, the direct detection constraint is expressed as follows:

\begin{equation} \label{pan1}
\frac{\sigma}{\text{M}_{\text{DM}}} \equiv \xi_S \frac{\sigma_s}{M_S} + \xi_{\psi} \frac{\sigma_{\psi}}{M_{\psi}} + \xi_V \frac{\sigma_v}{M_V} \lesssim \frac{\sigma}{M} \bigg\rvert_{\text{PandaX-4T}} \simeq 0.0005 \, \frac{\text{zb}}{\text{GeV}},
\end{equation}
Where $\xi_i$ is the dark matter fractions used to quantify the relative importance of each type of dark matter, as follows:

\begin{equation} \label{DMfraction}
\xi_{S}=\frac{\Omega_{S}}{\Omega_{DM}}, \;\;\; \xi_{\psi}=\frac{\Omega_{\psi}}{\Omega_{DM}}, \;\;\; \xi_{V}=\frac{\Omega_{V}}{\Omega_{DM}},
\end{equation}
where $\xi_S + \xi_\psi + \xi_V = 1$. Using dark matter relic density measurements from the Planck collaboration, we identify nine suitable points presented in Table \ref{tableOmega}. In this table, we have obtained the mass of each dark matter component, the vacuum expectation values $(\nu_2)$ for the ($\phi$) field, the allowed values for the coupling constant $\lambda_{\phi s}$ for the interaction between the scalar field (s) and ($\phi$), the mass of the second Higgs particle (dark Higgs), and the relic density values for the selected points.

\begin{table}[H] 
\centering
%\vspace{1cm}
\parbox{9cm}{\caption{Characteristics of the nine selected points. \label{tableOmega}}}

\vspace{7pt}
\begin{tabular}{ c c c c c c c c }
\hline 
Point&$M_V$(GeV)&$M_S$(GeV)&$M_{\psi}$(GeV)&${\nu}_2$(GeV)&$\lambda_{\phi s}$&$M_{H_2}$(GeV)&${\Omega}h^2$ \vspace{2pt} \\
\hline 
A&54.5&1125&611&567&8&125.79&0.12 \vspace{2pt} \\
\hline
B&330&665&175&340&8&124&0.14 \vspace{2pt} \\
\hline
C&300&725&220&400&8&125&0.14 \vspace{2pt} \\
\hline
D&800.2&50&600&290&$10^{-4}$&126.72&0.13 \vspace{2pt} \\
\hline
E&1125.25&1500&750.25&630&10&450.65&0.10 \vspace{2pt} \\
\hline
F&600.4&430&360.4&430&$10^{-2}$&120&0.97 \vspace{2pt} \\
\hline
G&500.5&287.5&300.5&390&1&84.52&0.96 \vspace{2pt} \\
\hline
H&1000.33&533.33&667&630&$10^{-4}$&203.31&0.97 \vspace{2pt} \\
\hline
I&334&50&200.67&310&$10^{-4}$&34&0.13 \vspace{2pt} \\
\hline
\end{tabular}
\end{table}
\medskip
Considering the condition $ M_V < 2 M_{\psi} $, all scalar, spinor, and vector components are viable DM candidates. Thus, Table \ref{tableOmega} shows that for each point, the mass of the vector particle is less than twice that of the spinor particle.

\begin{figure}[H] 
\begin{center}
\centerline{\hspace{0cm}\epsfig{figure=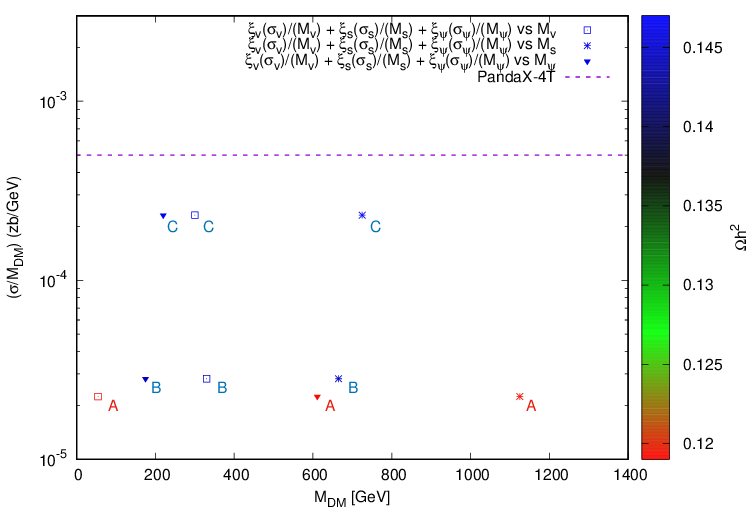,width=6.5cm}\hspace{0.5cm}\epsfig{figure=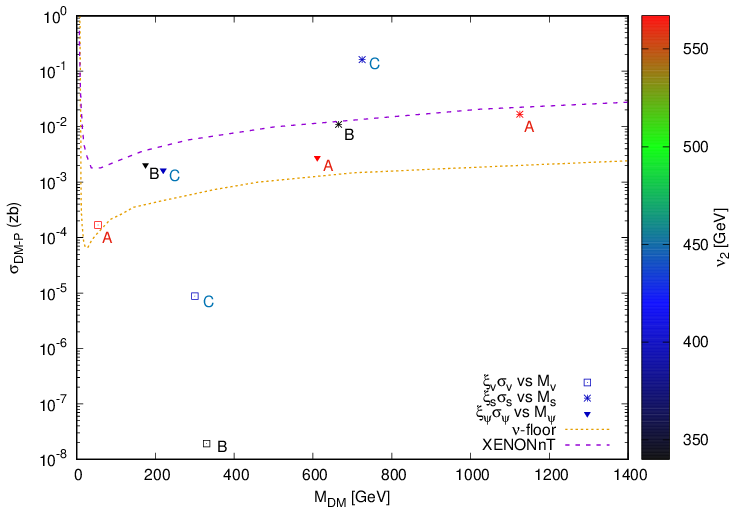,width=6.5cm}}
\centerline{\vspace{-0.7cm}}
\parbox{12cm}{\caption{These figures illustrate the points from table \ref{tableOmega} in relation to rescaled DM-Proton cross section versus DM masses, with $\nu_2$ and relic density ${\Omega}h^2$ represented by color bar. \label{pandagraph}}} 
\end{center} 
\end{figure}

The rescaled DM-nucleon cross sections of some selected points, relative to the threshold of the PandaX-4T and XENONnT detectors, are shown in Figure \ref{pandagraph}.

In the following sections, we will calculate gravitational waves from phase transitions using the points in table \ref{tableOmega}.

\section{Electroweak phase transition} \label{sec3}
\subsection{One-loop effective potential} \label{sec3-1}
As the gravitational waves from the electroweak phase transition obtain from the potential, in this section we acquire the effective potential as a function of temperature (T) and the scalar field ($\phi$). The effective potential is the combination of the zero-temperature and non-zero temperature potentials, including the tree-level and daisy potentials. In the flat direction, substituting $H_1=0$ and $H_2=\nu$  into equation (\ref{vtree}) results $V^{tree}=0$, as illustrated  in equation (\ref{hypotenuse}). The effective potential is defined as follows:

\begin{equation} \label{Vtotal1}
V_{\text{eff}}(\phi, T) = \underbrace{V^{\text{tree}}}_{0} + V_{(T=0)}^{(1-\text{Loop})} + V_{(T \neq 0)}^{(1-\text{Loop})} + V_{\text{daisy}}
\end{equation}

The effective potential was initially studied at the 1-loop level by Coleman and Weinberg. The Coleman–Weinberg potential is a sum of 1PI 1-loop diagrams featuring arbitrary numbers of external fields and particles in the loops \cite{Coleman:1973jx}. Then Gildener and Weinberg presented their formulation for a scale-invariant theory involving multiple scalar fields \cite{Gildener:1976ih}. The Gildener and Weinberg potential is as follows:

\begin{equation} \label{Vgw}
V_{\text{GW}}^{(1-\text{Loop})} = a \phi^4 + b \phi^4 \ln \left(\frac{\phi^2}{\Lambda^2}\right),
\end{equation}
where

\begin{align}  \label{Vfactor}
a &= \frac{1}{64 \pi^2 \nu^4} \sum_{k=1}^{n} g_k M_k^4 \left( \ln \frac{M_k^2}{\nu^2} - C_k \right), \nonumber \\
b &= \frac{1}{64 \pi^2 \nu^4} \sum_{k=1}^{n} g_k M_k^4.  
\end{align}  

In equation (\ref{Vgw}) $\Lambda$ is the renormalization group (RG) scale. In equation (\ref{Vfactor}) the $C_k$=3/2 (5/6), $M_k$ and $g_k$ are scalars/spinors (vectors), the measured mass of particles, and  the number of degrees of freedom of particle k, respectively. By finding the minimum of potential (\ref{Vgw}) at $\phi \neq 0$, the VEV value becomes non-zero:

\begin{align}  \label{Derivative}
\frac{dV^{1\text{-loop}}}{d\varphi} \bigg|_{\langle \varphi \rangle \neq 0} & = 0, \nonumber \\
\frac{d^{2}V^{1\text{-loop}}}{d\varphi^{2}} \bigg|_{\langle \varphi \rangle \neq 0} & > 0,  
\end{align} 
which leads to

\begin{equation}  \label{lambda}
\langle \varphi \rangle = \nu = \Lambda e^{-\left( \frac{a}{2b} + \frac{1}{4} \right)} \quad \text{and} \quad b > 0.  
\end{equation} 

By substituting (\ref{lambda}) and (\ref{Vfactor}) into equation (\ref{Vgw}), we obtain the effective potential at the $T=0$ as follows: 

\begin{equation}  \label{VT0}
V_{(T=0)}^{(1\text{-Loop})}(\varphi) = \frac{M_{H_2}^2}{8\nu^2}\varphi^4\left(\ln\left(\frac{\varphi^2}{\nu^2}\right) - \frac{1}{2}\right).
\end{equation}

We now examine the finite-temperature 1-loop effective potential, allowing us to compute the scalar field vacuum expectation values in a thermal bath at temperature T \cite{Dolan:1973qd}. The 1-loop finite-temperature corrections are given by

\begin{equation}  \label{VTN0}
V_{(T \neq 0)}^{(1\text{-Loop})}(\varphi, T) = \frac{T^4}{2\pi^2} \sum_{i=1}^{n} g_k J_{(B, F)_i}(x), \quad x = \frac{M_i \varphi}{\nu T},
\end{equation}
with thermal functions
\begin{equation} \label{JBF}
J_{\mathbf{B,F}}(x) = \int_{0}^{\infty} dy \, y^{2} \ln \left( 1 \mp e^{-\sqrt{ y^{2} + x^{2} }} \right).
\end{equation}
The $J_{(B_i)}(x)$ and $J_{(F_i)}(x)$  are simplified as follows \cite{Mohamadnejad:2019vzg}:

\begin{align}  \label{Termalfunction}
J_{(B_i)}(x) &= -\sum_{k=1}^{3} \frac{1}{k^2} x^2 k_2(kx), \nonumber \\
J_{(F_i)}(x) &= -\sum_{k=1}^{2} \frac{(-1)^k}{k^2} x^2 k_2(kx).  
\end{align}  

\medskip

Since we have only calculated the potential for one-loop Feynman diagrams so far, it is necessary to compute the potential for higher-order corrections of Feynman diagrams as well. For this purpose, we will use the daisy potential. We have obtained two types of potentials from the total effective potential, with the daisy potential being the last component. The daisy potential includes thermal contributions from virtual particles at finite temperatures, modifying the effective potential by incorporating thermal excitations. Daisy diagrams serve as higher loop corrections that capture the effects of thermal fluctuations and interactions among virtual particles, making the effective potential temperature-dependent, especially during phase transitions. This non-perturbative effect reflects significant feedback from these loop corrections, leading to important physical insights that traditional perturbative approaches might miss. 
The daisy potential is defined as follows \cite{Carrington:1991hz}:

\begin{equation}  \label{Vdaisy}
V_{\text{daisy}}(\varphi, T) = \sum_{k=1}^{n} \frac{g_k T^4}{12 \pi} \left( x^3 - \left(x^2 + \frac{\prod_k}{T^2}\right)^{\frac{3}{2}} \right), \quad x = \frac{M_i \varphi}{\nu T}, 
\end{equation} 
where the sum includes only scalar bosons and the longitudinal degrees of freedom of the gauge bosons. In equation (\ref{Vdaisy}), the $\prod_k$ are the thermal masses of particle k, defined as follows:

\begin{align}  \label{Daisytermal}
\Pi_s &= \frac{T^2}{24} \left( \lambda_{HS} + 6\lambda_S + \lambda_{\phi S} \right), \nonumber \\
\Pi_w &= \frac{11}{6} (g_{SM}^2 T^2), \quad g_{SM} = \frac{2M_w}{\nu_1}, \nonumber \\
\Pi_V &= \frac{2}{3} (g_V^2 T^2), \quad g_V = \frac{M_V}{\nu_2}, \nonumber \\
\Pi_Z &= \frac{11}{6} (g_{SM}^2 T^2), \nonumber \\
\Pi_{H_1} &= \frac{T^2}{24} \left( \frac{9}{2} g_{SM}^2 + \frac{3}{2} g_{SM}^{\prime 2} + 6\lambda_t^2 + \lambda_H \phi + 6\lambda_H + \lambda_{HS} \right), \quad g_{SM}' = \frac{2\sqrt{M_Z^2 - M_w^2}}{\nu_1}, \nonumber \\
\lambda_t &= \frac{\sqrt{2} M_t}{\nu_1},
\end{align}
where $\lambda_t$ denotes the top quark Yukawa coupling, and $g_v$, $g_{SM}$, and $g_{SM}^\prime$ are dark $U(1)_D$, $SU(2)_L$, and $U(1)_Y$ gauge couplings, respectively.
The parameters values from (\ref{constrins}) are utilized to calculate the thermal masses in (\ref{Daisytermal}).
Finally, we can express the total effective potential (\ref{Vtotal1}) in terms of temperature (T) and the scalar field ($\phi$). We can now substitute the effective potential into the Euclidean action. The function $S_3(T)$ represents the three-dimensional Euclidean action for a spherically symmetric bubble, defined as follows:

\begin{equation} \label{S3} 
S_3(T) = 4\pi \int_0^{\infty} dr \, r^2 \left( \frac{1}{2} \left( \frac{d\varphi}{dr} \right)^2 + V_{\text{eff}}(\varphi, T) \right),  
\end{equation} 
where $\phi$ fulfills the differential equation that minimizes $S_3(T)$:

\begin{equation}  \label{S3derive}
\frac{d^2 \varphi}{dr^2} + \frac{2 d\varphi}{r dr} = \frac{\partial V_{\text{eff}}(\varphi, T)}{\partial \varphi},  
\end{equation}  
with the following boundary conditions:

\begin{equation}  \label{S3conditions}
\frac{d\varphi}{dr} \bigg|_{r=0} = 0, \quad \text{and} \quad \varphi(r \to \infty) = 0.  
\end{equation} 

\medskip

\begin{figure}[H] 
\centering
  \begin{minipage}{0.32\textwidth}
    \centering
    \includegraphics[width=\linewidth]{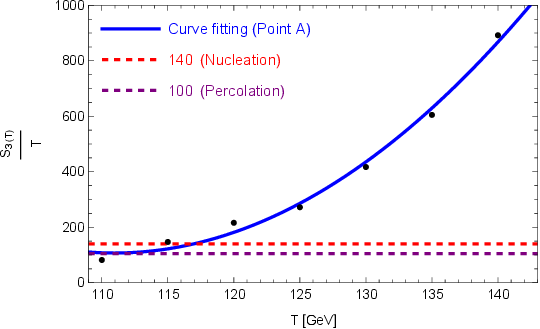}
  \end{minipage}
  \hfill
  \begin{minipage}{0.32\textwidth}
    \centering
    \includegraphics[width=\linewidth]{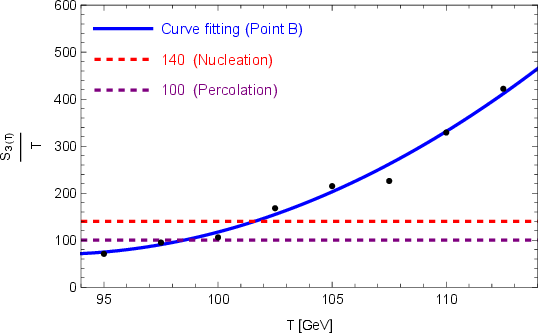}
  \end{minipage}
  \hfill
  \begin{minipage}{0.32\textwidth}
    \centering
    \includegraphics[width=\linewidth]{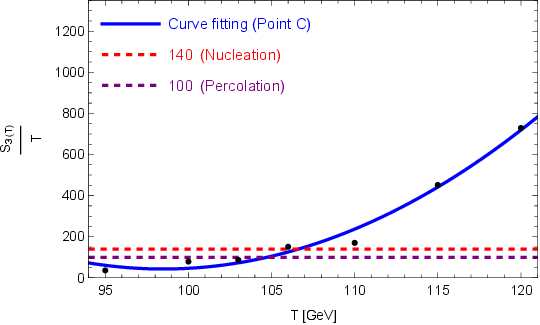}
  \end{minipage}
  
  \vspace{0cm} % ########

  \begin{minipage}{0.32\textwidth}
    \centering
    \includegraphics[width=\linewidth]{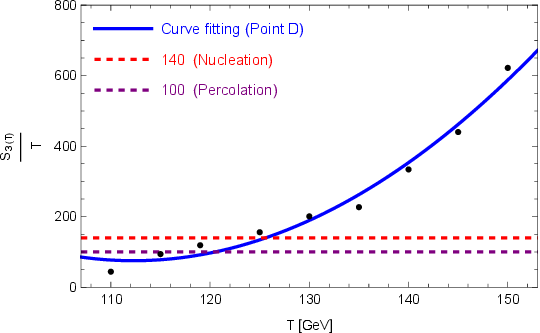}
  \end{minipage}
  \hfill
  \begin{minipage}{0.32\textwidth}
    \centering
    \includegraphics[width=\linewidth]{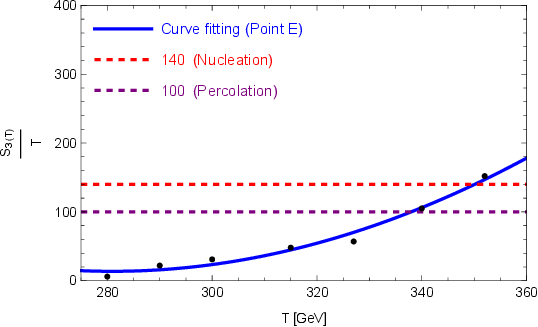}
  \end{minipage}
  \hfill
  \begin{minipage}{0.32\textwidth}
    \centering
    \includegraphics[width=\linewidth]{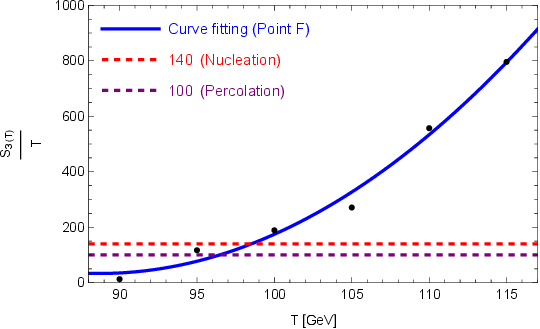}
  \end{minipage}
  
  \vspace{0cm} % ########

  \begin{minipage}{0.32\textwidth}
    \centering
    \includegraphics[width=\linewidth]{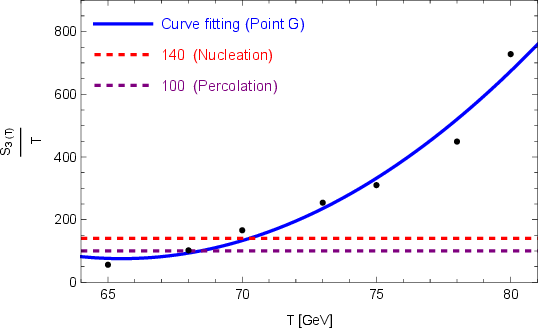}
  \end{minipage}
  \hfill
  \begin{minipage}{0.32\textwidth}
    \centering
    \includegraphics[width=\linewidth]{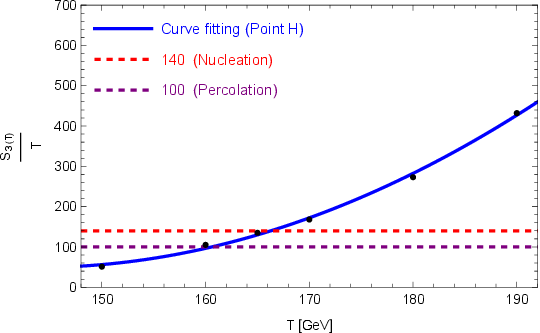}
  \end{minipage}
  \hfill
    \begin{minipage}{0.32\textwidth}
    \centering
    \includegraphics[width=\linewidth]{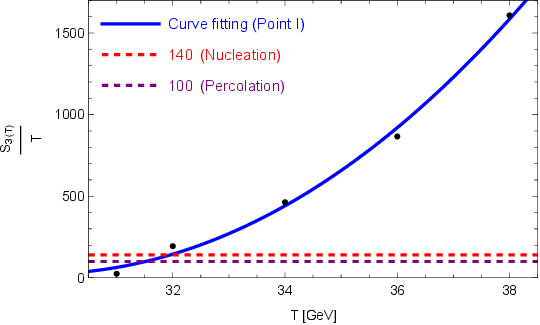}
  \end{minipage}
  
\parbox{12.5cm}{\caption{The blue line shows $S_3(T)/T$ as a function of 
T, while the dashed red and purple lines indicate the values $S_3(T)/T=140$ and 
$S_3(T)/T=100$, respectively, at which nucleation and percolation occur. The black dots are derived using the AnyBubble package. \label{graphAction-T}}} 
\end{figure}

The nucleation temperature ($T_n$) and percolation temperature ($T_p$) occur where the corresponding euclidean action are $S_n = \frac{S_3(T_n)}{T_n} \simeq 140$ \cite{Apreda:2001us,Athron:2023xlk} and $S_p = \frac{S_3(T_p)}{T_p} \simeq 100$ \cite{Caprini:2019egz}, respectively. To solve equation (\ref{S3derive}) and find the Euclidean action (\ref{S3}), we used the AnyBubble package \cite{Masoumi:2017trx}. This package allows us to calculate the Euclidean action for each selected point using the two minima in the effective potential diagram as function of the scalar field ($\phi$), and plot the ratio of the Euclidean action to temperature (T). In figure \ref{graphAction-T}, we present $\frac{S_3(T)}{T}$ for selected points in table \ref{tableOmega}. 

\begin{figure}[H] 
\centering
  \begin{minipage}{0.32\textwidth}
    \centering
    \includegraphics[width=\linewidth]{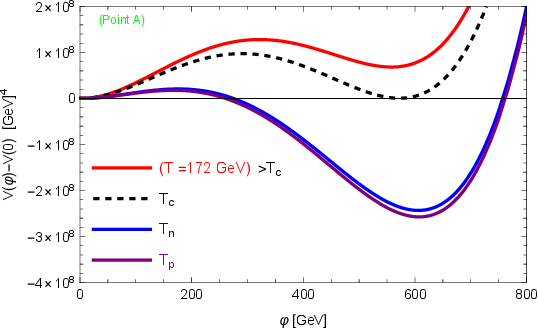}
  \end{minipage}
  \hfill
  \begin{minipage}{0.32\textwidth}
    \centering
    \includegraphics[width=\linewidth]{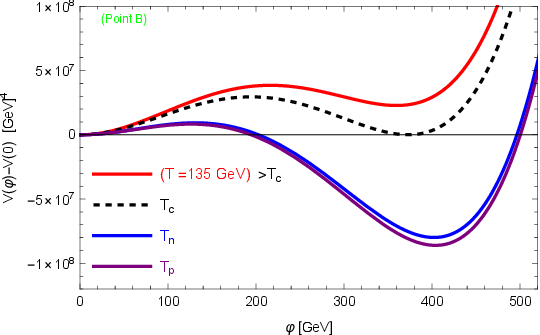}
  \end{minipage}
  \hfill
  \begin{minipage}{0.32\textwidth}
    \centering
    \includegraphics[width=\linewidth]{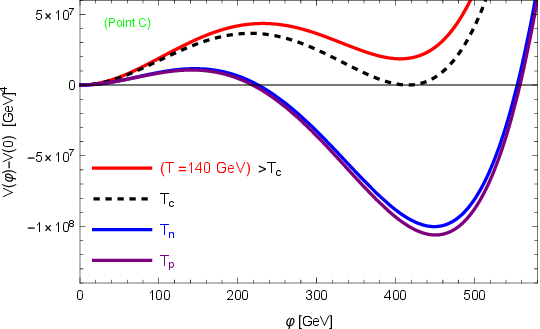}
  \end{minipage}
  
  \vspace{0cm} % ########

  \begin{minipage}{0.32\textwidth}
    \centering
    \includegraphics[width=\linewidth]{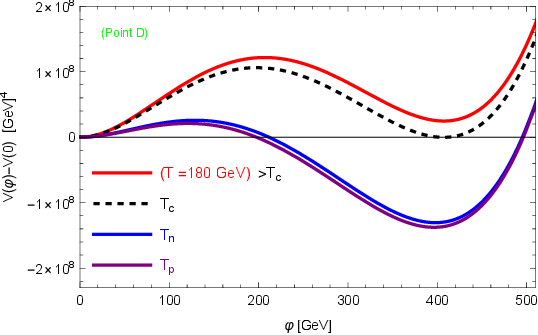}
  \end{minipage}
  \hfill
  \begin{minipage}{0.32\textwidth}
    \centering
    \includegraphics[width=\linewidth]{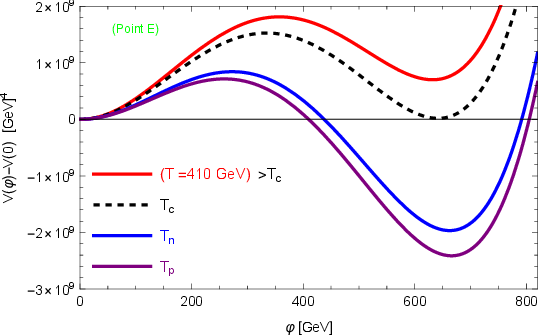}
  \end{minipage}
  \hfill
  \begin{minipage}{0.32\textwidth}
    \centering
    \includegraphics[width=\linewidth]{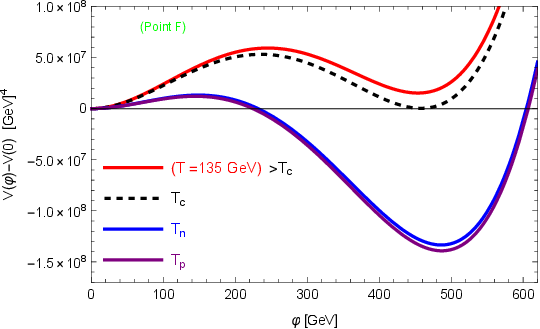}
  \end{minipage}
  
  \vspace{0cm} % ########

  \begin{minipage}{0.32\textwidth}
    \centering
    \includegraphics[width=\linewidth]{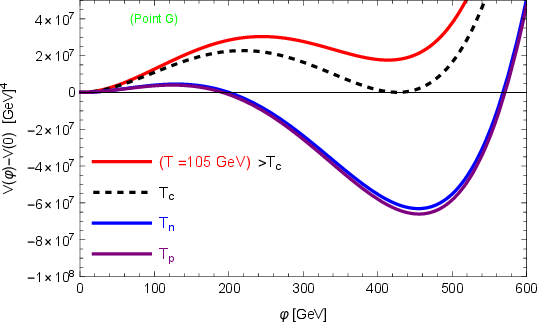}
  \end{minipage}
  \hfill
  \begin{minipage}{0.32\textwidth}
    \centering
    \includegraphics[width=\linewidth]{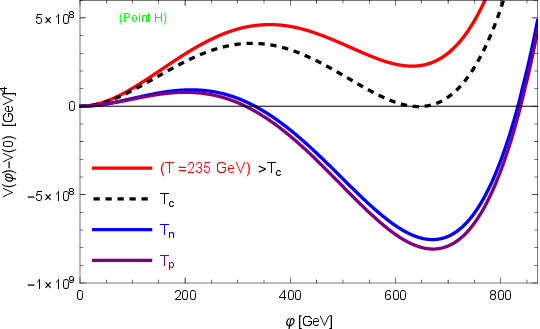}
  \end{minipage}
  \hfill
    \begin{minipage}{0.32\textwidth}
    \centering
    \includegraphics[width=\linewidth]{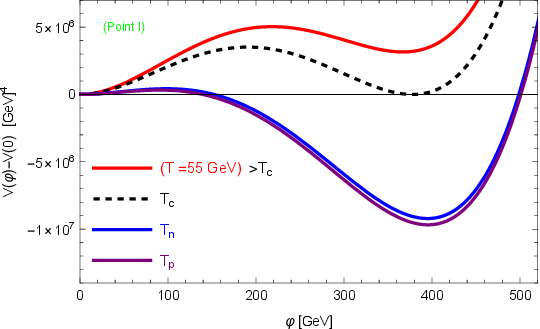}
  \end{minipage} 
\parbox{12.5cm}{\caption{Effective potential diagrams of points in table \ref{tableTC} as a function of $\phi$ scalar field at percolation ,nucleation, critical, and supercritical temperatures. \label{graphV_PHI}}}    
\end{figure}

In Figure \ref{graphAction-T}, the nucleation temperature ($T_n$) and percolation temperature ($T_p$) are determined by finding the intersection points of the curve fitting and the horizontal lines. To determine the critical temperature $T_c$ for the points in table \ref{tableOmega}, we must plot the effective potential (\ref{Vtotal1}) against the scalar field ($\phi$) to identify the temperature at which the effective potential exhibits two aligned minima (see figure \ref{graphV_PHI}).

\noindent Figure \ref{graphV_PHI} shows the effective potential for each point in Table \ref{tableOmega} at the four temperatures: percolation ($T_p$), nucleation ($T_n$), critical ($T_c$), and supercritical ($T>T_c$). To ensure \( V_{\text{eff}}(0, T) = 0 \) at all temperatures, we subtract a constant term from the potential: \( V_{\text{eff}}(\phi, T) \rightarrow V_{\text{eff}}(\phi, T) - V_{\text{eff}}(0, T) \). As shown in Figure \ref{graphV_PHI}, at the nucleation temperature ($T_n$), the zero minimum is known as a false vacuum and the negative minimum is known as a true vacuum. There is a bulge between the two minima, which prevents the electroweak phase transition from occurring smoothly between the minima, and instead occurs suddenly and explosively, like a dam breaking. 

\noindent The supercritical temperatures ($T>T_c$) shown in Figure \ref{graphV_PHI} are determined as arbitrary and indicate the position of the effective potential in relation to the percolation, nucleation and critical temperatures.
Figure \ref{graphV_PHI} illustrates that as the temperature decreased, the effective potential also diminishes. Also, for all selected points, the thermal order is as follows:
\begin{equation} \label{orderT}
T_c>T_n>T_p.
\end{equation}

As the separation between the critical and nucleation temperatures grows, more vacuum energy is released during the false-to-true vacuum transition. This regime is known as supercooling. The released energy can temporarily heat the primordial plasma, a process called reheating, but it is not enough to raise the negative minimum in Figure \ref{graphV_PHI} above the zero minimum and to create a reverse phase transition. During reheating, the higher temperature suppresses the formation of new bubbles, effectively increasing the spacing between bubble nuclei and allowing existing bubbles to grow larger before colliding. Larger bubble collisions produce stronger gravitational waves. Thus, supercooling enhances reheating, which in turn strengthens the gravitational-wave signal.

The difference between the critical and nucleation temperatures determines the degree of supercooling. Therefore, the supercooling parameter is as follows:

\begin{equation} \label{supercooling}
\delta_{sc,n}=\frac{T_c-T_n}{T_c}.
\end{equation}

\noindent When bubbles formation occurs along with supercooling, the released heat reaches a level where it can heat significant the initial plasma. As the system cools, the plasma temperature returns to the nucleation temperature. This time delay gives true vacuum bubbles enough time to occupy about 30\% of the universe's volume. Therefore, in the temperature versus $\delta_{sc,n}$ plot, for $\delta_{sc,n} \gtrsim 0.5$, the percolation-temperature curve tends to a constant value and no longer decreases \cite{Athron:2022mmm}. In Table \ref{tableTC}, the supercooling parameter $\delta_{sc,n}$ is calculated for each selected points.

\begin{table}[H] 
\centering
%\vspace{1cm}
\parbox{12cm}{\caption{Thermal characteristics of the selected points in Table \ref{tableOmega}. \label{tableTC}}}

\vspace{7pt}
\begin{tabular}{ c c c c c c c }
\hline 
Point&$T_p$(GeV)&$T_n$(GeV)&$T_C$(GeV)&$[{\phi}_{V_{min}}]_{T_p}$(GeV)&$[{\phi}_{V_{min}}]_{T_n}$(GeV)&$\delta_{sc,n}$ \vspace{2pt} \\
\hline 
A&112&116&162.5&607.471&606.022&0.28 \vspace{2pt} \\
\hline
B&98.5&101&128.2&404.004&402.363&0.21 \vspace{2pt} \\
\hline
C&104.7&106.7&135.3&450.592&449.134&0.21 \vspace{2pt} \\
\hline
D&120&126&174.5&396.343&397.913&0.27 \vspace{2pt} \\
\hline
E&338&350&396&665.93&662.441&0.11 \vspace{2pt} \\
\hline
F&96&98&132&486.955&486.125&0.25 \vspace{2pt} \\
\hline
G&68&70&99&455.481&454.619&0.29 \vspace{2pt} \\
\hline
H&160&166&222.5&671.402&670.249&0.25 \vspace{2pt} \\
\hline
I&31.5&32&51&395.154&395.05&0.37 \vspace{2pt} \\
\hline
\end{tabular}
\end{table}

\noindent In Table \ref{tableTC}, the $[{\phi}_{V_{min}}]_{T_n}$ denotes the point of ${\phi}$ scalar field at which the effective potential value in $T_n$ is minimized, as illustrated in Figure \ref{graphV_PHI}. In this context, the $[{\phi}_{V_{min}}]_{T_n}$ represents the true vacuum. According to the values in the $\delta_{sc,n}$ column of Table \ref{tableTC}, supercooling ($\delta_{sc,n} \gtrsim 0.5$) has not occurred for the selected points in our model.

In the next section, we will use the values from Table \ref{tableTC} to calculate the background gravitational waves.

\subsection{Gravitational waves} \label{sec3-2}
The background gravitational waves from strong first-order electroweak phase transitions has three contributions:
\begin{enumerate}
\item The collision of spherical bubble walls in the plasma after the Big Bang.
\item Sound waves that release kinetic energy in the plasma due to bubble collisions.
\item Turbulence of massive plasma in the fabric of space-time.
\end{enumerate}

Before calculating gravitational waves, we need to introduce $\alpha$ (the strength parameter) and $\beta$, the inverse duration of the phase transition, as follows:

\begin{align}  \label{alphaGW}
& \alpha = \frac{\Delta \left( V_{\text{eff}} - T \frac{\partial V_{\text{eff}}}{\partial T} \right)_{\big|_{T_p}}}{\rho_*}, \;\;\; \rho_* = \frac{\pi^2 g_*}{30} T_p^4 \nonumber \\
& \frac{\beta}{H_*} = T_p \left. \frac{d}{dT} \left( \frac{S_3(T)}{T} \right) \right|_{T_p},
\end{align}
where $g_*=100$. In equation (\ref{alphaGW}), we obtain the parameter $\frac{\beta}{H_*}$ from the slope of the graphs in figure \ref{graphAction-T} at the intersection point of the curve fitting and the purple line. Now, we derive the density relations of gravitational waves for all three types of their sources, including collision (coll), sound wave (sw), and turbulence (turb) \cite{Athron:2024xrh}. We begin with the density of background gravitational waves from collision sources \cite{Huber:2008hg,Jinno:2016vai}:
\begin{equation}  \label{omegacoll}
\Omega_{\text{coll}}^{env}(f)h^2 = \quad 1.67 \times 10^{-5} \, \Delta \left(  \frac{100}{g_*} \right)^{\frac{1}{3}} \left( \frac{H_*}{\beta} \right)^2  \left( \frac{\kappa_{\phi} \alpha}{1 + \alpha} \right)^2 S_{\mathrm{env}}(f / f_{\mathrm{env}})  ,  
\end{equation} 
with
\begin{equation} \label{delta1}
\Delta = \frac{0.48 v_{w}^3}{1 + 5.3 v_{w}^2 + 5 v_{w}^4},
\end{equation}
$S_{\mathrm{env}}(r)$ is the spectral shape parameter of the bubble, defined by the following relation:
\begin{equation} \label{spherical}
S_{\mathrm{env}}(r) = \left( 0.064 r^{-3} + 0.456 r^{-1} + 0.48 r \right)^{-1},
\end{equation}
where $r={f}/{f_{env}}$, and $f$ is the GW frequency of the gravitational wave, with peak frequency at:
\begin{align}  
& \frac{f_{\mathrm{env}}}{1\, \mathrm{\mu Hz}} = 16.5 \left( \frac{f_*}{\beta} \right) \left( \frac{g_*}{100} \right)^{\frac{1}{6}} \left( \frac{\beta}{H_*} \right) \left( \frac{T_*}{100\, \mathrm{GeV}} \right), \nonumber \\ 
& \frac{f_*}{\beta} = \frac{0.35}{1 + 0.069 v_{w} + 0.69 v_{w}^4}. \label{peak}
\end{align}
In equation (\ref{omegacoll}), $\kappa_{\phi}$ is the fraction of latent heat converted into the gradient energy of the Higgs-like field, defined as follows \cite{Kamionkowski:1993fg}:
\begin{equation}  \label{Kcoll}
\kappa_{\phi} = \frac{1}{1 + 0.715 \alpha} \left[ 0.715 \alpha + \frac{4}{27} \sqrt{\frac{3 \alpha}{2}} \right],  
\end{equation} 
and $v_w$ is the bubble propagation speed \cite{Bodeker:2009qy}. 

\medskip
\medskip
\noindent The density of background GW from sound wave sources is defined as \cite{Hindmarsh:2015qta,Hindmarsh:2013xza}:

\begin{equation}  \label{Omegasw}
\Omega_{\text{sw}}(f)h^2 = 2.061 \, F_{gw,0} \, \Gamma^2 \, \bar{U}_f^4 \, S_{sw}(f) \, \tilde{\Omega}_{gw} \min \left( H_* R_* / \bar{U}_f, 1 \right) \left( H_* R_* \right) h^2,
\end{equation}
where
\begin{align} 
& F_{gw,0} = 3.57 \times 10^{-5} \left( \frac{100}{g_*} \right)^{\frac{1}{3}}, \nonumber \\
& R_{*} = \left( 8\pi \right)^{1/3} \frac{v_{w}}{\beta}, \nonumber \\
& K = \frac{\kappa_{sw} \alpha}{1 + \alpha} = \Gamma \bar{U}_f^2, \label{gammaU}
\end{align}
the $R_{*}$ and $K$ are bubble separation and kinetic energy fraction, respectively.

\noindent In equation (\ref{Omegasw}), the $\Gamma={4}/{3}$ and $\tilde{\Omega}_{gw}\sim0.012$ are ratio of enthalpy to the energy density and determined from simulations, respectively.
Also, for fluid the spectral shape parameter $S_{\text{sw}}(f)$ of the bubble, defined as follows \cite{Hindmarsh:2017gnf,Caprini:2019egz}:

\begin{align}
S_{sw}(f) &= \left( \frac{f}{f_{sw}} \right)^3 \left( \frac{7}{4 + 3(f / f_{sw})^2} \right)^\frac{7}{2}, \nonumber \\
\frac{f_{sw}}{1\, \mu \mathrm{Hz}} &= 2.6 \left( \frac{z_{p}}{10} \right) \left( \frac{T_*}{100\, \mathrm{GeV}} \right) \left( \frac{g_*}{100} \right)^{\frac{1}{6}} \left( \frac{1}{H_* R_*} \right), \;\;\; z_{p}\sim10. \label{Fsw}
\end{align}
the $\kappa_{sw}$ in equation (\ref{gammaU}) is determined as \cite{Espinosa:2010hh}:

\begin{equation} \label{Ksw}
\kappa_{sw} = 
\begin{cases}
\displaystyle \frac{ c_s^{11/5} \kappa_A \kappa_B }{ (c_s^{11/5} - v_w^{11/5})\kappa_B + {v_w}{c_s}^{6/5} \kappa_A }, & v_w \leq c_s \\
\kappa_B + (v_w - c_s) \delta \kappa + \frac{(v_w - c_s)^3}{(v_J - c_s)^3}l_k , & c_s < v_w < v_J \\
\displaystyle \frac{(v_J - 1)^3 v_{J}^{5/2} v_{w}^{-5/2}\kappa_C \kappa_D}{[(v_J - 1)^3-(v_w - 1)^3]v_J^{5/2}\kappa_C+(v_w - 1)^3\kappa_D}, & v_J \leq v_w
\end{cases}
\end{equation}
where
\begin{align}
\kappa_A &\simeq v_{w}^{6/5} \frac{6.9 \alpha}{1.36 - 0.037 \sqrt{\alpha} + \alpha}, \nonumber \\
\kappa_B &\simeq \frac{\alpha^{2/5}}{0.017 + (0.997 + \alpha)^{2/5}}, \nonumber \\
\kappa_C &\simeq \frac{\sqrt{\alpha}}{0.135 + \sqrt{0.98 + \alpha}}, \nonumber \\
\kappa_D &\simeq \frac{\alpha}{0.73 + 0.083 \sqrt{\alpha} + \alpha}, \nonumber \\
\delta \kappa &\simeq -0.9 \log \frac{\sqrt{\alpha}}{1 + \sqrt{\alpha}}, \nonumber \\
l_\kappa &\simeq \kappa_C - \kappa_B - (v_J - c_s) \delta \kappa, \nonumber \\
v_J &= \frac{1}{1 + \alpha} \left( c_s + \sqrt{\alpha^2 + \frac{2 \alpha}{3}} \right). \label{Vj}
\end{align}
The density of background GW from turbulence sources is defined as \cite{Caprini:2009yp,Binetruy:2012ze}:
\begin{equation}  \label{Omegaturb}
\Omega_{\text{turb}}(f) h^2= 3.35 \times 10^{-4} v_w \left( \frac{H_*}{\beta} \right) \left( \frac{\kappa_{\text{turb}} \alpha}{1 + \alpha} \right)^{3/2} \left( \frac{100}{g_*} \right)^{1/3} \frac{r^3}{(1 + r)^{11/3} (1 + 8 \pi f / H_0)}
\end{equation}
where $r={f}/{f_{turb}}$ and $\kappa_{turb}$ is the magneto-hydrodynamic (MHD) turbulence, defined as follows \cite{Hindmarsh:2015qta}:

\begin{equation}  \label{Kturb}
\kappa_{turb} \simeq (0.05 \;\; to \;\; 0.1)\times \kappa_{sw}.
\end{equation}
In equation (\ref{Omegaturb}), the peak frequency $f_{\text{turb}}$ and the red-shifted Hubble rate at GW generation $H_0$ are defined as follows:

\begin{align}
& f_\text{turb} = {27} \frac{1}{v_w} \left( \frac{\beta}{H_n} \right) \left( \frac{T_*}{100\, \text{GeV}} \right) \left( \frac{g_*}{100} \right)^{1/6} \mu \text{Hz}, \nonumber \\
& H_0 = 16.5 \left( \frac{g_*}{100} \right)^{\frac{1}{6}} \left( \frac{T_*}{100\, \text{GeV}} \right) \,\mu \text{Hz} . \label{Fturb}
\end{align}

In the next section, we plot all types of background gravitational wave densities obtained in terms of GW frequency. 

\section{Results} \label{sec4}
In this section, by using the previous section’s relations and the results of table \ref{tableBetaH}, we can determine the total background gravitational wave density from the electroweak phase transition as follows:

\begin{equation} \label{Omegatotal} 
\Omega_{\text{total}}(f) h^2 \simeq  \Omega_{\text{coll}}^{env}(f)h^2 +  \Omega_{\text{sw}}(f) h^2 + \Omega_{\text{turb}}(f) h^2. 
\end{equation} 
The values of $\alpha$ as the strength parameter and $\frac{\beta}{H_*}$ as the inverse duration of the phase transition, for selected points, as shown in Table \ref{tableBetaH}.

\begin{table}[H] 
\centering
%\vspace{1cm}
\parbox{10cm}{\caption{The background gravitational wave parameters resulting from the phase transition, obtained using (\ref{alphaGW}), correspond to the points listed in Table \ref{tableOmega}. \label{tableBetaH}}}

\vspace{7pt}
\begin{tabular}{ c c c c c}
\hline 
Point&$\alpha$&$\frac{\beta}{H_*}$&${\Omega h^2}_{max}^{(total)}$&$f_{{\Omega h^2}_{max}^{(total)}} \; (HZ)$ \vspace{2pt} \\
\hline 
A&0.123291&138.71&$8.63\times10^{-13}$&$1.42\times10^{-4}$ \vspace{2pt} \\
\hline
B&0.104421&1008.06&$1.06\times10^{-14}$&$8.71\times10^{-4}$ \vspace{2pt} \\
\hline
C&0.104247&1903.97&$2.95\times10^{-15}$&$1.77\times10^{-3}$ \vspace{2pt} \\
\hline
D&0.0386934&676.682&$1.51\times10^{-15}$&$7.4\times10^{-4}$ \vspace{2pt} \\
\hline
E&0.0338373&1012.39&$4.57\times10^{-16}$&$3.17\times10^{-3}$ \vspace{2pt} \\
\hline
F&0.145771&1551.5&$1.06\times10^{-14}$&$1.28\times10^{-3}$ \vspace{2pt} \\
\hline
G&0.232053&965.032&$8.56\times10^{-14}$&$5.49\times10^{-4}$ \vspace{2pt} \\
\hline
H&0.100534&929.561&$1.12\times10^{-14}$&$1.34\times10^{-3}$ \vspace{2pt} \\
\hline
I&0.524421&2561.93&$6.78\times10^{-14}$&$7.27\times10^{-4}$ \vspace{2pt} \\
\hline
\end{tabular}
\end{table}

\begin{figure}[H]
\centering
  \begin{minipage}{0.32\textwidth}
    \centering
    \includegraphics[width=\linewidth]{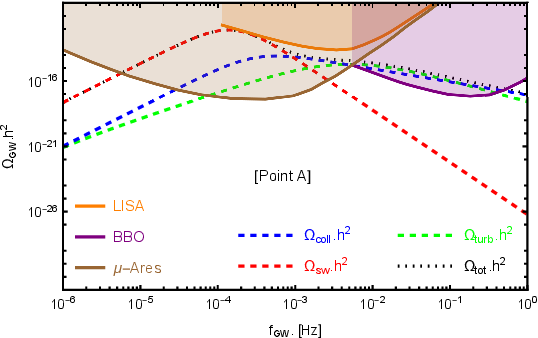}
  \end{minipage}
  \hfill
  \begin{minipage}{0.32\textwidth}
    \centering
    \includegraphics[width=\linewidth]{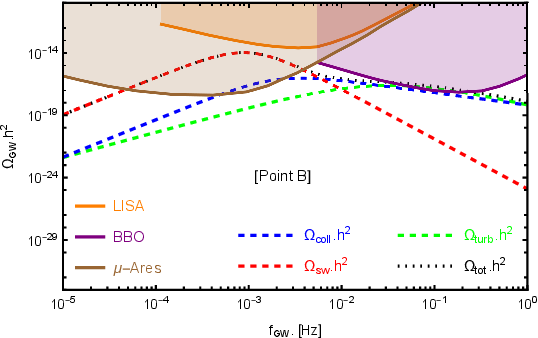}
  \end{minipage}
  \hfill
  \begin{minipage}{0.32\textwidth}
    \centering
    \includegraphics[width=\linewidth]{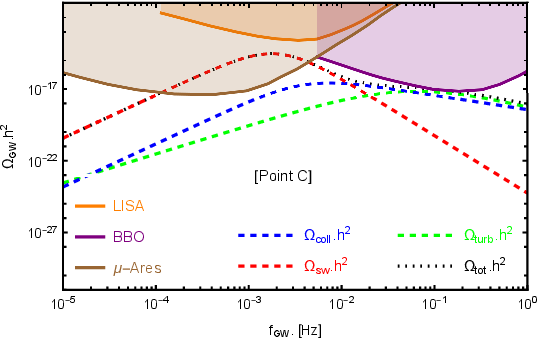}
  \end{minipage}
  
  \vspace{0cm} % ########

  \begin{minipage}{0.32\textwidth}
    \centering
    \includegraphics[width=\linewidth]{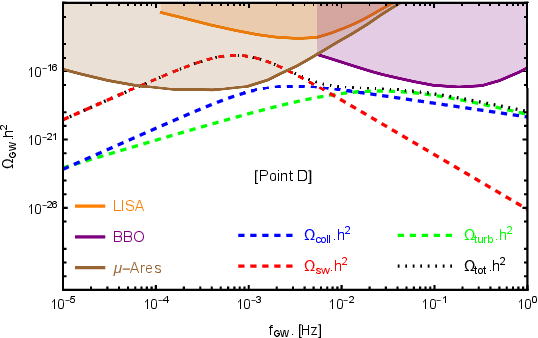}
  \end{minipage}
  \hfill
  \begin{minipage}{0.32\textwidth}
    \centering
    \includegraphics[width=\linewidth]{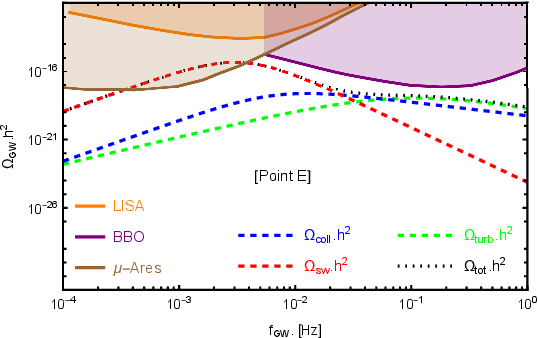}
  \end{minipage}
  \hfill
  \begin{minipage}{0.32\textwidth}
    \centering
    \includegraphics[width=\linewidth]{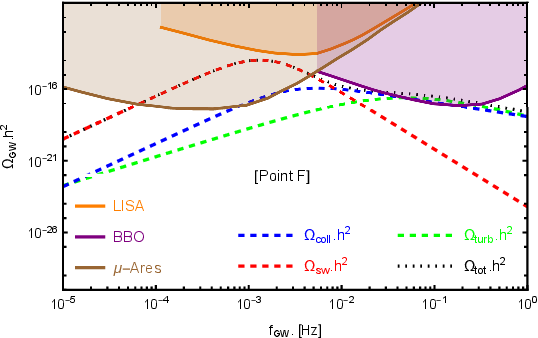}
  \end{minipage}
  
  \vspace{0cm} % ########

  \begin{minipage}{0.32\textwidth}
    \centering
    \includegraphics[width=\linewidth]{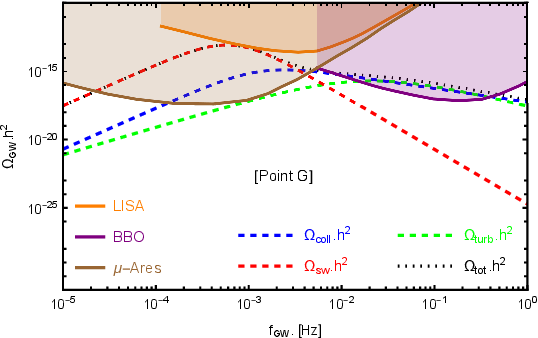}
  \end{minipage}
  \hfill
  \begin{minipage}{0.32\textwidth}
    \centering
    \includegraphics[width=\linewidth]{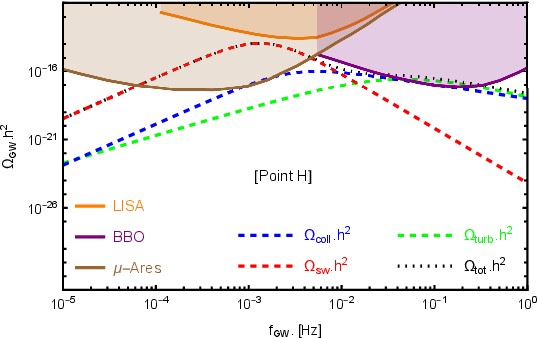}
  \end{minipage}
  \hfill
    \begin{minipage}{0.32\textwidth}
    \centering
    \includegraphics[width=\linewidth]{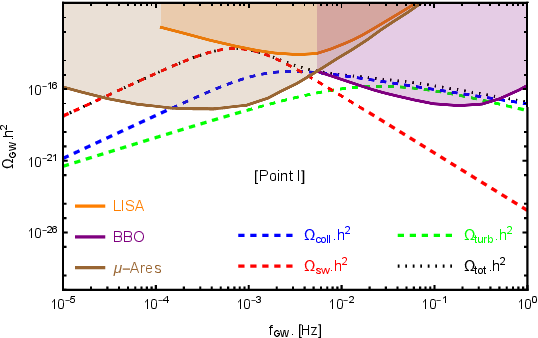}
  \end{minipage}
  \parbox{12.5cm}{\caption{GW spectrum as a function of backgound gravitational frequency for points in Table \ref{tableBetaH}, considering $v_w=1$. \label{graphFrequency}}} 
\end{figure}

Finally, figure \ref{graphFrequency} presents the total background gravitational wave density (\ref{Omegatotal}) versus GW frequency, utilizing data from table \ref{tableBetaH} for selected points.

\noindent In Figure \ref{graphFrequency}, there are three curves for Lisa, BBO and $\mu$-Ares, which show the power of space-based laser detectors that have not yet been launched into space. As shown in Figure \ref{graphFrequency}, the LISA detector has limited capability to detect the background gravitational waves produced by the electroweak phase transition for the selected points in parameter space. In contrast, the BBO and $\mu$-Ares detectors can detect a portion of these background gravitational waves for all selected points.

\noindent The background gravitational wave density from sound wave sources, depicted in all graphs of figure \ref{graphFrequency}, is the primary contributor to the total gravitational wave background at low frequencies. However, as the frequency increases, the contribution from turbulence and collision become dominant. The frequency of approximately $10^{-3}$ Hz marks the maximum values in most plots and delineates the optimal frequency range for detecting the gravitational wave background within our model. As the bubble propagation speed rises from zero to one, the gravitational wave density initially increases, reaching its maximum at around $v_w=0.7$, and then decreases thereafter.

\begin{figure}[H]
\centering
  \begin{minipage}{0.32\textwidth}
    \centering
    \includegraphics[width=\linewidth]{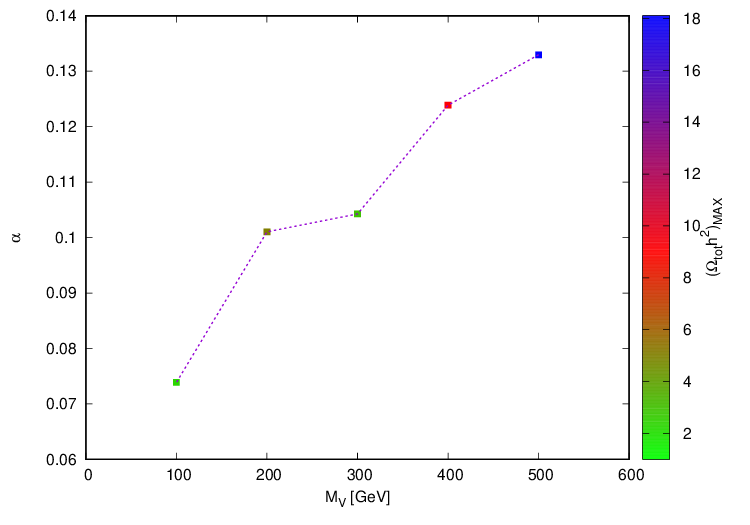}
  \end{minipage}
  \hfill
  \begin{minipage}{0.32\textwidth}
    \centering
    \includegraphics[width=\linewidth]{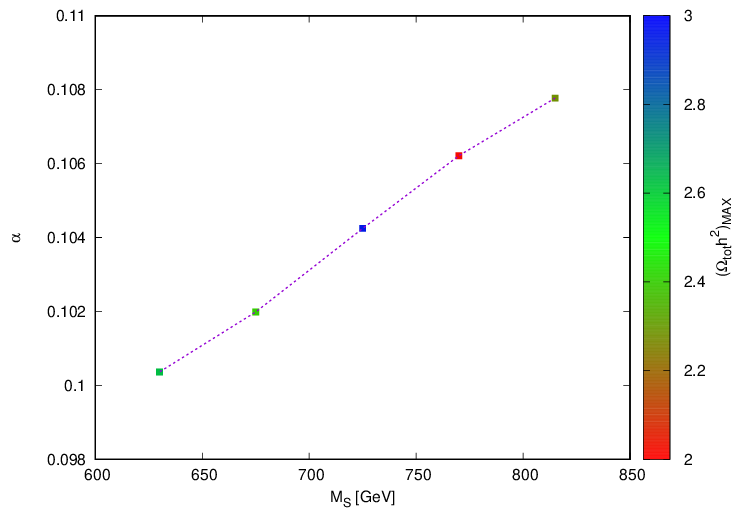}
  \end{minipage}
  \begin{minipage}{0.32\textwidth}
    \centering
    \includegraphics[width=\linewidth]{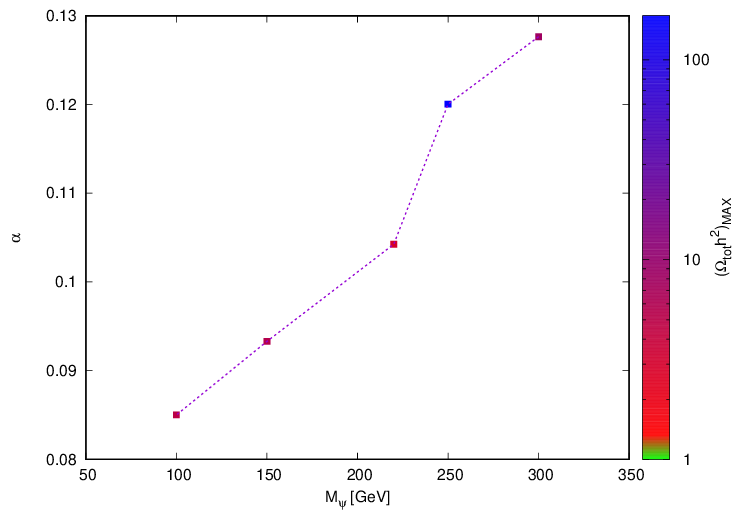}
  \end{minipage}
  
  \vspace{0cm} % ########
    \centering
  \begin{minipage}{0.32\textwidth}
    \centering
    \includegraphics[width=\linewidth]{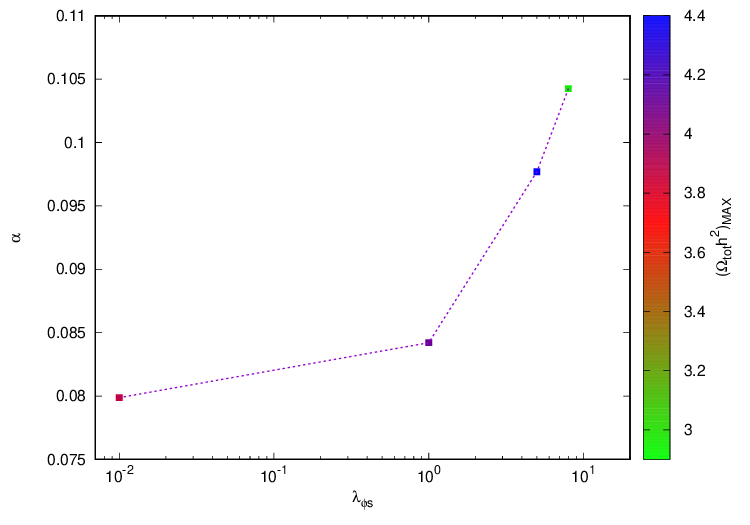}
  \end{minipage}
    \begin{minipage}{0.32\textwidth}
    \centering
    \includegraphics[width=\linewidth]{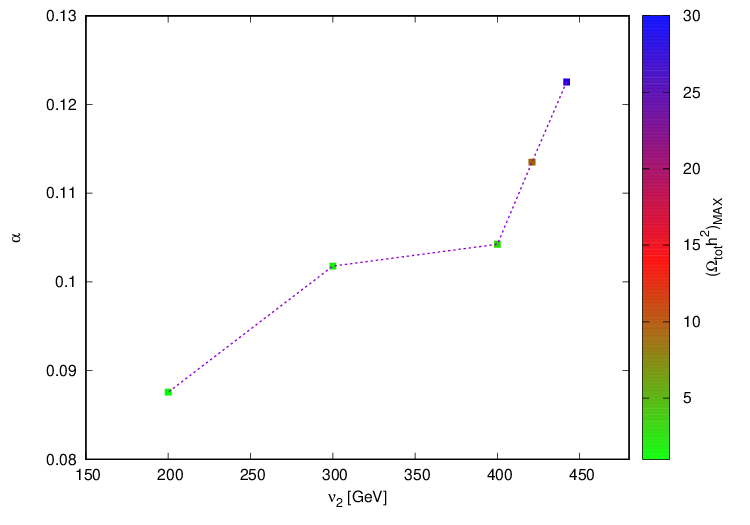}
  \end{minipage}
\parbox{8.5cm}{\caption{Variations in $\alpha$ as a function of changes in each independent parameter around point C in parameter space, with $({\frac{{\Omega}h^2}{10^{-15}}})$ represented by color bar.  \label{AlphaCc}}}
\end{figure}

\begin{figure}[H]
\centering
  \begin{minipage}{0.32\textwidth}
    \centering
    \includegraphics[width=\linewidth]{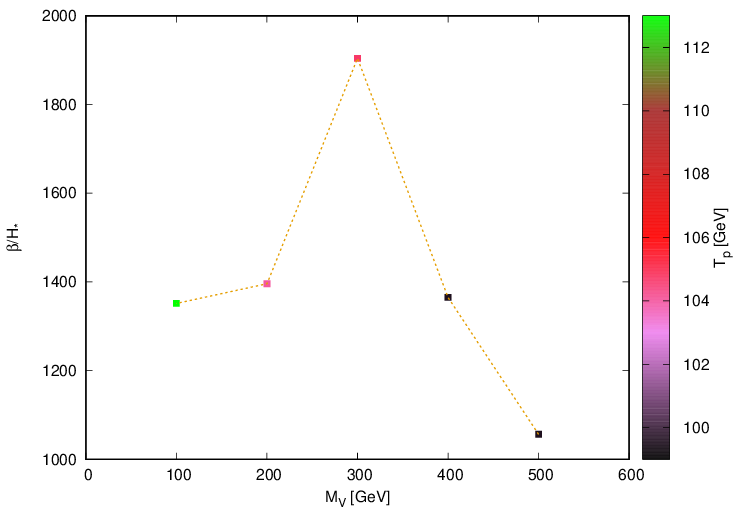}
  \end{minipage}
  \hfill
  \begin{minipage}{0.32\textwidth}
    \centering
    \includegraphics[width=\linewidth]{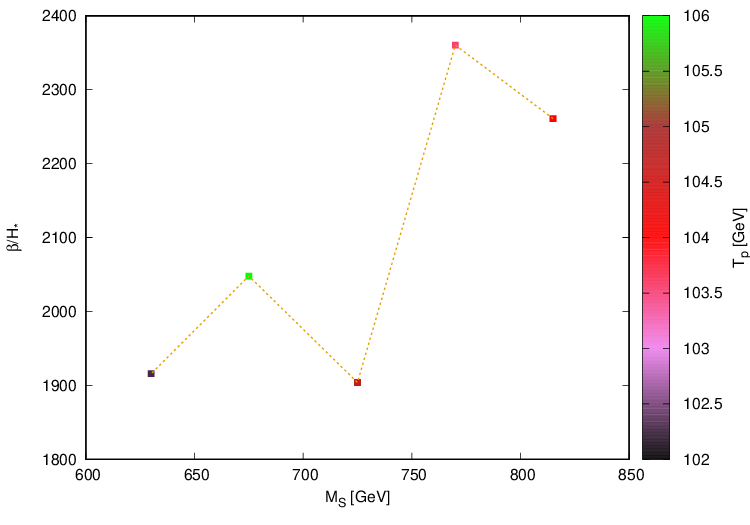}
  \end{minipage}
  \begin{minipage}{0.32\textwidth}
    \centering
    \includegraphics[width=\linewidth]{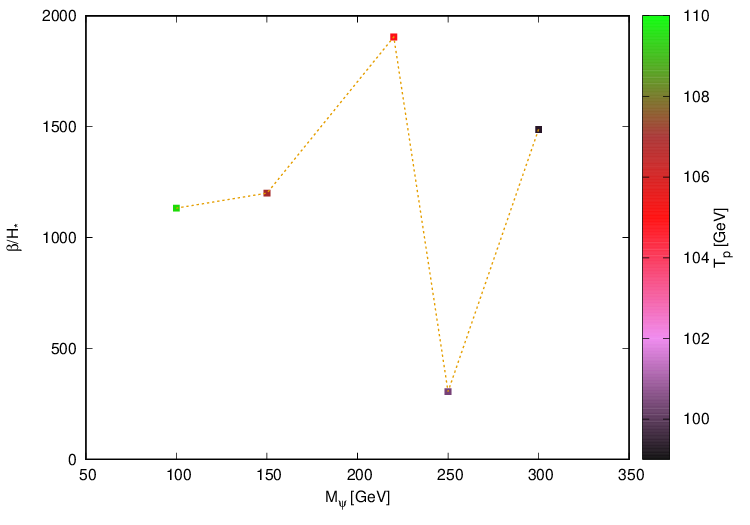}
  \end{minipage}
  
  \vspace{0cm} % ########

  \begin{minipage}{0.32\textwidth}
    \centering
    \includegraphics[width=\linewidth]{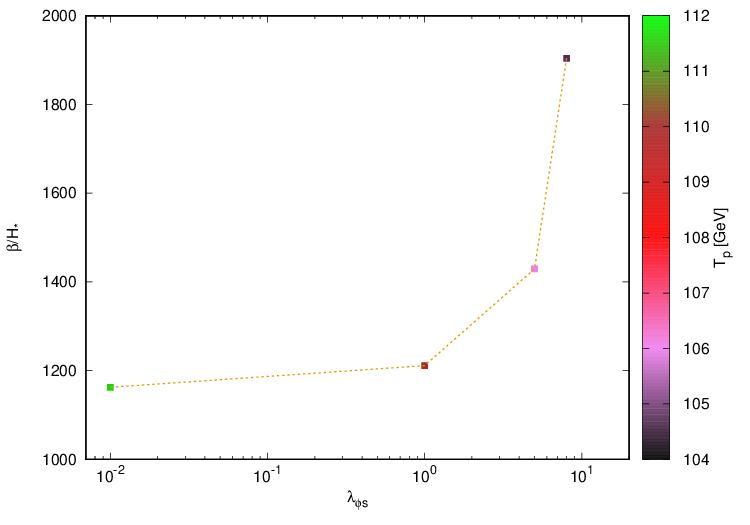}
  \end{minipage} 
  \begin{minipage}{0.32\textwidth}
    \centering
    \includegraphics[width=\linewidth]{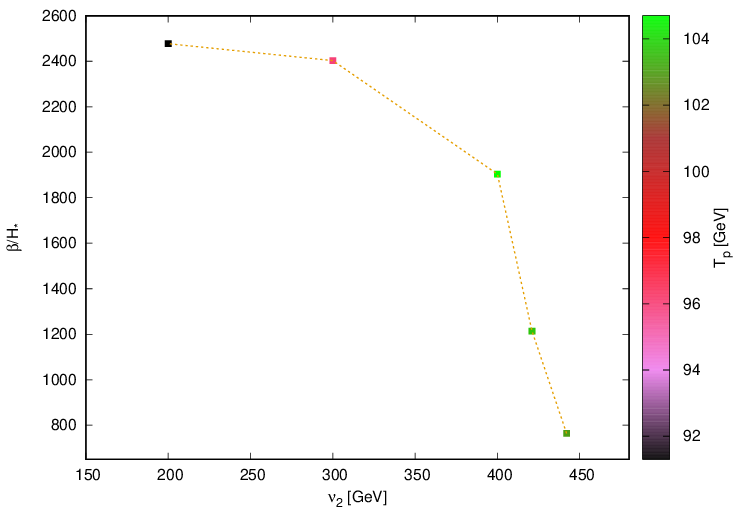}
  \end{minipage}
  \parbox{8.5cm}{\caption{Variations in $\frac{\beta}{H_*}$ as a function of changes in each independent parameter around point C in parameter space, with $T_p$ represented by color bar. \label{BetaCc}}} 
\end{figure}

In order to understand the effect of changes in the mass of each dark matter component and other independent parameters of the model on gravitational waves, we have changed the values of the parameters around point C and plotted their effects on the values of $\alpha$ and $\frac{\beta}{H_*}$ in the graphs of Figures \ref{AlphaCc} and \ref{BetaCc}.

\noindent As shown in Figure \ref{AlphaCc}, increasing the independent dark matter parameters leads to a monotonic rise in $\alpha$ value.

\noindent In Figures \ref{AlphaCc} and \ref{BetaCc}, we held four parameters constant in each run and varied a single remaining parameter to determine the corresponding gravitational wave parameters.

\noindent According to the graphs in Figure \ref{BetaCc}, the value of $\frac{\beta}{H_*}$ has no significant relationship with the values of dark matter mass, but it has a direct relationship with ${\lambda}_{\phi s}$ and an inverse relationship with ${\nu}_2$.

\section{Conclusion} \label{sec5}
This paper revisits a model of dark matter in which particles exhibit three spin types: a vector particle, a scalar, and a spinor. In our model, the Higgs particle acts as a portal between Standard Model particles and dark matter. We identified some points in parameter space that exceed the detection capability of PandaX-4T and XENONnT, while their relic density matches the Planck satellite's findings. We examined the electroweak phase transition for selected points and showed that a potential barrier between the two minima of the effective potential at nucleation temperature can generate gravitational waves. The three parameters, $\alpha$, $\frac{\beta}{H_*}$, and percolation temperature ($T_p$), are sufficient for gravitational wave calculations, independent of other model parameters. Considering the values in Table \ref{tableBetaH} and the revised background gravitational wave density calculation for each selected point, we find that the ${\Omega h^2}_{GW}$ increases with the $\alpha$ value and decreases with the $\frac{\beta}{H_*}$ value (i.e., a direct relationship with $\alpha$ and an inverse relationship with $\frac{\beta}{H_*}$). Our results indicate that the sensitivity of the LISA detector is insufficient for detecting cosmic background gravitational waves for all benchmarks of Table \ref{tableOmega}, whereas the BBO and $\mu$-Ares detectors have the capability to confirm the existence of these waves.

\providecommand{\href}[2]{#2}\begingroup\raggedright\endgroup

\end{document}